\documentclass[a4paper, twocolumn,showpacs,prl,aps,10pt]{revtex4-1}
\usepackage{graphicx} 
\usepackage{subfigure}
\usepackage{amsmath}
\usepackage{amssymb}
\usepackage{wasysym}
\usepackage{dcolumn}
\usepackage{tikz}
\usepackage{float}
\usepackage{graphicx}
\usepackage{epstopdf}
\usepackage{amsmath, amsthm, amssymb, esint}
\usepackage{graphicx}
\usepackage{physics}
\usepackage{booktabs}
\usepackage{siunitx} 
\usepackage{gensymb}
\usepackage{indentfirst}
\usepackage[english]{babel}
\usepackage{multirow}
\usepackage{mathtools}
\usepackage{tikz}
\usepackage{pgfplots}

\begin{document}
\DeclareGraphicsExtensions{.pdf,.png,.jpg,.eps}

\title{Size and Shape of Condensing Steam Jets and Plumes}
\author{N. Fourcade$^{1}$, F. Poydenot$^{1}$, P. Langlois$^{1}$, F. Barbe$^{1}$, N. Sinnathamby$^{1}$, P. Claudin$^{2}$ and B. Andreotti$^{1}$}
\affiliation{
$^{1}$Laboratoire de Physique de l'Ecole Normale Sup\'erieure, UMR 8023 PSL-SU-UPCit\'e-CNRS, Paris, France\\
$^{2}$Physique et M\'ecanique des Milieux H\'et\'erog\`enes, UMR 7636 ESPCI-PSL-SU-UPCit\'e-CNRS, Paris, France}
 
\date{9 October 2026}
\begin{abstract}
Droplet-laden cloud dynamics emerge from coupled phase transitions and turbulent mixing. Combining experiments and mean-field theory, we establish scaling laws for steam-injected clouds in unsaturated air seeded with condensation nuclei. Depending on whether the cloud is confined to the initial momentum-dominated zone of the jet or extends into the buoyancy-dominated plume, two asymptotic regimes arise. In the momentum regime, cloud dimensions scale with the injector radius. In the buoyancy regime, size follows a \(\dot{m}^{2/5}\) scaling, where \(\dot{m}\) is the injected steam mass flow rate. Assuming interdroplet air remains humidity saturated, our mean-field analysis predicts cloud persistence governed by enthalpy and humidity mixing. These findings bridge microphysical processes and macroscale dynamics, offering a quantitative framework for atmospheric cloud evolution.
\end{abstract}

\maketitle

Dense sprays and aerosols are utilized in medicine~\cite{prather_airborne_2020,bourouiba_fluid_2021,chong_extended_2021,poydenot_risk_2022}, industry~\cite{villermaux_fragmentation_2007}, and agriculture~\cite{hewitt_spray_2000}. They also govern mass and heat transfers in natural systems, notably in cloud formation. Cloud condensation nuclei (CCNs) seed water droplet formation~\cite{erinin_droplet_2025} and originate from aerosolized terrestrial and marine particles~\cite{veron_ocean_2015}. Clouds exert a dual climate feedback. They cool the planet by reflecting solar radiation and increasing Earth's albedo. Simultaneously, they entrain water vapor to higher altitudes, where infrared absorption enhances the greenhouse effect~\cite{stevens_atmospheric_2005,bony_clouds_2015}. This energy balance depends on cloud organization and dynamics~\cite{de_rooy_entrainment_2013}, raising the question: what determines the evaporation altitude of clouds following their buoyant ascent?

From a fluid dynamics perspective, warm clouds are multiphase systems where suspended droplets interact with turbulent airflow. The life cycle of cloud droplets, including nucleation, phase changes, and collisional aggregation, is governed by the turbulent mixing of water vapor, enthalpy, and CCNs. The multiscale and intermittent nature of turbulence~\cite{bodenschatz_can_2010} induces spatiotemporal variability in these processes. Conversely, microphysical transformations influence the flow. Latent heat release modifies buoyancy, droplet radiative properties affect energy budgets, and drag forces alter airflow dynamics~\cite{poydenot_gap_2024,baker_effects_1984,squires_entraining_1962}. This two-way coupling between turbulence and microphysics governs cloud dynamics and climate impact.
\begin{figure}[t!]
\includegraphics{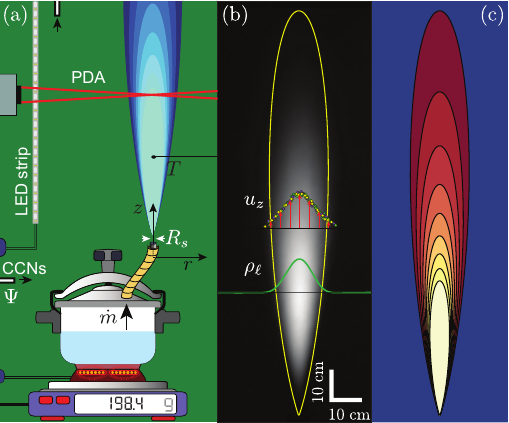}
\vspace{-4 mm} 
\caption{(a) Schematic representation of the experimental setup. (b) Visualization of the cloud generated using an injector with radius $R_s = 1.5~\text{mm}$ and a vapor mass flow rate $\dot{m} = 0.55~\text{g/s}$. The superimposed transverse profiles illustrate the diffused light intensity and axial velocity. The yellow contour delineates the cloud boundary predicted by the mean-field model ($\rho_{\ell} = 0$). (c) Isocontours of the similarity field $\phi(r,z)$, as defined by Eq.~\eqref{formphi}.}
\vspace{-4 mm}
\label{fig:Fig1}
\end{figure}

\begin{figure*}[t!]
\includegraphics{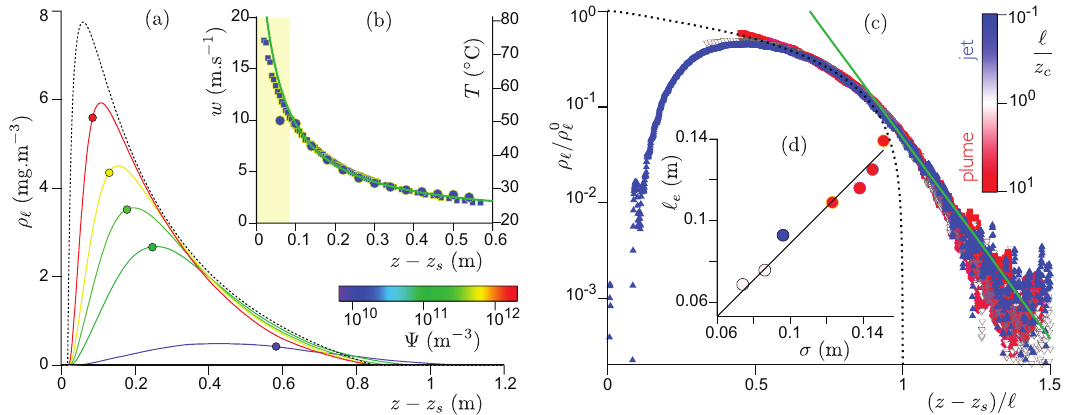}
\vspace{-4 mm}
\caption{(a) Axial profiles of liquid density $\rho_\ell$ as a function of altitude $z-z_s$ for a filtered atmosphere ($\Psi=6.4\times 10^9\;{\rm m^{-3}}$, violet) and with increasing amounts of incense smoke up to $\Psi=1.3\times 10^{12}\;{\rm m^{-3}}$ (red). The dotted line shows the asymptotic model prediction assuming instantaneous phase change. The symbols show the predicted cross-over altitude $z_d$. (b) Axial velocity $w$ (circles, left axis) and temperature $T$ (squares, right axis) profiles for $R_s = 1~\text{mm}$ and $\dot{m} = 0.15~\text{g/s}$. The green solid line corresponds to the $1/z$ decay predicted by the self-similar jet model [Eq.~\eqref{wjet}], with a fitted virtual origin $z_s$, consistently found $\approx R_s/\lambda$. Yellow region: near-field premixing zone where the asymptotic regime is not yet established. (c) Centerline liquid density normalized by $\rho_\ell^0$, the model density extrapolated at the outlet ($z=z_s$), as a function of the rescaled altitude $(z-z_s)/\ell$, in log-linear scale, for all runs. Symbol colors code for the ratio $\ell/z_c$ (color bar), distinguishing clouds confined to the jet regime ($\ell/z_c \ll 1$, blue) from those extending into the plume regime ($\ell/z_c \gg 1$, red). Dotted line: model prediction as in (a), vanishing at $z=\ell$. Green line: exponential fit of the decay in the upper cloud region, defining the decay length $\ell_e$. (d) This length is displayed as a function of the maximum cloud width $\sigma$ (evaluated at $z \approx \ell$). The solid line indicates the proportional relationship $\ell_e \propto \sigma$.}
\vspace{-4 mm}
\label{fig:Fig2}
\end{figure*}

To investigate these mechanisms, we conduct a laboratory experiment characterizing the droplet-laden region (the ``cloud'') generated by injecting a steam jet into unsaturated air containing CCNs~\cite{baskaya_radial_1997,lesniewski_particle_1998,oerlemans_experimental_2001,garmory_aerosol_2008,campbell_turbulent_2010,di_veroli_modeling_2011,zhou_simulation_2014,pesmazoglou_aerosol_2014,lim_understanding_2018}. After a spatial transient that shortens as CCN concentration $\Psi$ increases, evaporation and condensation occur rapidly relative to the flow time. Leveraging similarity solutions for the jet flow~\cite{morton_turbulent_1956,schwarz_radial_1963,list_turbulent_1982,lee_micro-droplet_2022}, we predict the cloud's spatial extent under this rapid phase-change assumption. The cloud size exhibits two asymptotic scalings. It is independent of the steam mass flow rate \(\dot m\) when confined to the momentum-dominated zone, and independent of the injector radius \(R_s\) when extending into the buoyancy-dominated zone.

\textit{Experimental setup}---The setup, shown in Fig.~\ref{fig:Fig1}(a), uses a sealed pressure cooker as a water vapor source. A controlled heater maintains a constant mass flow rate $\dot{m}$, measured by a digital scale. Vapor flows through a heated pipe to prevent premature condensation and exits a cylindrical nozzle of radius $R_s$. The exiting superheated vapor ($T_s \simeq 106~^\circ$C) has velocity $U_s$ and density $\rho_s$, yielding $\dot{m} = \rho_s \pi R_s^2 U_s$. A platinum resistance thermometer monitors the vapor temperature. A phase Doppler anemometer (PDA) measures droplet velocities and size distributions (diameters 0.5 to 22 $\mu$m)~\cite{albrecht_laser_2011}. This technique achieves low measurement uncertainty due to the steady mean flow and 10\% fluctuation rate. The macroscopic liquid density $\rho_\ell$ is the sum of individual droplet contributions. Droplet Lagrangian dynamics are statistically independent of their size $d$ if the Stokes number $\text{St} = \tau_S / \mathcal{T} \ll 1$. Here, $\tau_S = \rho_{\rm H_2O} d^2 / (18\eta)$ is the viscous response time (with air viscosity $\eta$ and water density $\rho_{\rm H_2O}$) and $\mathcal{T}$ is the flow's Lagrangian integral time \cite{poydenot_risk_2022}. Our PDA measurements confirm this expectation, showing no correlation between droplet velocity and diameter [see Supplemental Material (SM) \cite{Note1}]. Consequently, droplets and vapor share the same turbulent mixing rate, behaving as a single phase. Backlit imaging (15 s exposure) with uniform LED illumination visualizes the entire cloud. Since droplet diameters exceed the optical wavelength, the scattered light intensity follows Mie scattering and is proportional to the total droplet cross-sectional area ($\propto n d^2$, with number density $n$). Backscattering calibration is detailed in SM \cite{Note1}. Although dense cloud cores with high CCN concentrations induce multiple scattering, the cloud contour remains optically dilute, enabling robust edge detection. The experiment is housed in a 100 m$^3$ climatic chamber with controlled temperature $T_\infty$ and water vapor density $\rho_v^\infty$.  To systematically vary the background CCN concentration $\Psi$, the chamber air is conditioned either by filtering the ambient air or by burning incense sticks. The concentration $\Psi$, estimated as the maximum droplet density along the jet axis, scales linearly with the optically measured concentration of $2.5~\mu\text{m}$ suspended particles (PM2.5 standard).

\textit{Flow structure}---Figure~\ref{fig:Fig2}(a) displays the centerline liquid water density $\rho_\ell$ for various CCN concentrations $\Psi$. The nozzle injects slightly superheated vapor. Downstream, turbulent mixing of enthalpy, water vapor, and CCNs with ambient air proceeds at comparable rates. Because the saturation vapor density $\rho_{\mathrm{sat}}(T)$ depends exponentially on temperature, it drops faster than the total water content during mixing. This creates a supersaturated environment that drives condensation onto CCNs. Consequently, $\rho_\ell$ initially increases along $z$, defining a condensation zone that shortens as CCN concentration increases. The profiles then collapse onto a master decay curve, matching our mean-field model derived for saturated interstitial air (detailed below). Farther downstream, as the jet entrains dry air, conditions approach the unsaturated ambient state and evaporation dominates. These cloud droplets persist longer than isolated ones because collective evaporation maintains near saturation in the interstitial air. A logarithmic plot of $\rho_\ell$ [Fig.~\ref{fig:Fig2}(c)] reveals a transition from an initially linear decay to an exponential regime, a signature of heterogeneous turbulent mixing \cite{gollub_fluctuations_1991,villermaux_mixing_2003,de_rivas_dense_2016}.

\textit{Characteristic times}---The absence of intrinsic geometric length scales in this localized injection configuration promotes self-similar flow development. The mean vertical velocity profile, $u_z(r,z)$, exhibits a Gaussian transverse distribution with a characteristic width $\sigma = \lambda z$, where the spreading rate $\lambda = 0.07$ is constant across the explored parameter space [Figs.~\ref{fig:Fig1}(b) and S1 \cite{Note1}]. The virtual source is located at $z = 0$, a vertical distance $z_s$ below the actual injector outlet. The local turbulent mixing timescale is $\tau_{\mathrm{mix}} \sim \sigma/w$, where $w(z) = u_z(0,z)$ is the centerline axial velocity. Droplet growth over a characteristic time $\tau$ follows the diffusion-limited Maxwell-Langmuir scaling law $d^2/\tau \sim D(\rho_{v}-\rho_{\mathrm{sat}})/\rho_{\mathrm{H}_2\mathrm{O}}$, where $d$ is the droplet diameter and $D$ is the molecular diffusion coefficient of water vapor in air. The macroscopic liquid water density scales as $\rho_{\ell} \sim \Psi \rho_{\mathrm{H}_2\mathrm{O}} d^3$, where $\Psi$ is the background CCN concentration. The crossover between the near-injector region, where the total water content, of density $\rho_w = \rho_v + \rho_\ell$, is primarily vapor ($\rho_{v} \simeq \rho_w$), and the downstream regime, where the vapor saturates the interstitial space ($\rho_\ell \simeq \rho_{w}-\rho_{\mathrm{sat}}$), occurs when the condensed water mass equals the available excess vapor: $(\rho_{w}-\rho_{\mathrm{sat}}) \sim \Psi \rho_{\mathrm{H}_2\mathrm{O}} d^3$. This yields the scaling laws for $d$ and $\tau$:
\begin{equation}
d \sim \left( \frac{\rho_w - \rho_{\mathrm{sat}}}{\Psi \rho_{\mathrm{H}_2\mathrm{O}}} \right)^{1/3}
\! \text{and} \,\,\,
\tau \sim \frac{1}{D \Psi^{2/3}} \! \left( \frac{\rho_{\mathrm{H}_2\mathrm{O}}}{\rho_w - \rho_{\mathrm{sat}}} \right)^{1/3} .
\end{equation}
This demonstrates that the transient condensation zone shortens at higher CCN concentrations $\Psi$. The end of this region occurs when this microphysical relaxation time becomes comparable to the local advective mixing time. Equating these timescales defines the crossover distance $z_d$, which is determined in practice by numerically solving $\tau = 4\sigma/w$, accounting for the prefactors. Figure~\ref{fig:Fig2}(a) shows agreement between the derived scaling law and the experimental data, confirming this interpretation. Beyond $z_d$, the separation of timescales and a centerline $\rho_\ell$ distribution independent of $\Psi$ justify the assumption that the interstitial air maintains water vapor saturation.

\begin{figure}[t!]
\includegraphics{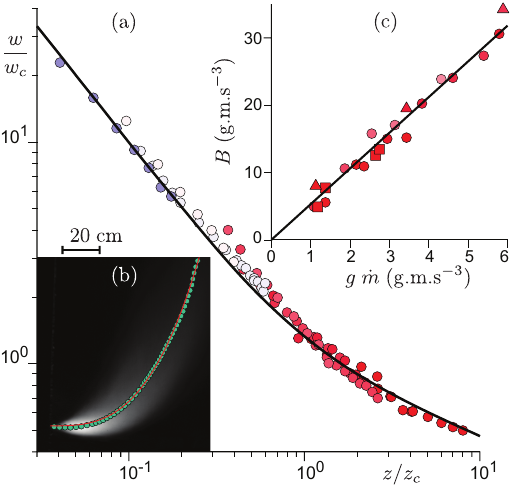}
\vspace{-4 mm}
\caption{(a) Centerline axial velocity profile: all data collapse when rescaled by jet-plume crossover values $w_c$ and $z_c$ [Eq.~(S4) of the SM \cite{Note1}]. Solid line: analytical function from Eq.~(S3) \cite{Note1}. (b),(c) Measurement of the effective buoyancy flux $B$. (b) Cloud formed by horizontal vapor injection (experimental conditions for this run: $R_s = 4.5~\text{mm}$, $\dot{m} = 0.22~\text{g/s}$). Green dots: central streamline. Red line: theoretical fit $z \propto B x^3$ [Eq.~(S9) of SM \cite{Note1}]. (c) $B$ vs mass flow rate $\dot{m}$ for various outlet radii. Circles (triangles in the presence of incense smoke) denote geometric measurements as in panel (b); squares indicate values deduced from the velocity data collapse (a). Solid line: linear fit $B = \beta g \dot{m}$, with $\beta = 5.3$. Symbol colors in (a) and (c) code for $\ell/z_c$, as in Fig.~\ref{fig:Fig2}(c).}
\vspace{-4 mm}
\label{fig:Fig3}
\end{figure}

\textit{Self-similar asymptotics}---Under noncondensing conditions, the hydrodynamic fields transition between two asymptotic regimes: a jet regime, where the momentum flux ($\rho_s \pi R_s^2 U_s^2$) is conserved, and a plume regime, where it grows monotonically with $z$ driven by a constant buoyancy flux $B = \int_0^\infty (\rho_\infty - \rho) g u_z \, 2\pi r \, \mathrm{d}r$~\cite{turner_jets_1966,sanchez_development_1989, Note1}. The centerline axial velocity scales as

\begin{eqnarray}
w(z) &=& \sqrt{\frac{\rho_s}{\rho_\infty}} \frac{R_s U_s}{\lambda z} \quad \text{(jet regime)}, \label{wjet} \\
w(z) &=& \left(\frac{3 B}{2 \pi \rho_\infty \lambda^2 z}\right)^{1/3} \quad \text{(plume regime)}, \label{wplume}
\end{eqnarray}

where $\rho_\infty$ is the ambient air density. All measured axial velocity profiles collapse onto a master curve when rescaled by $w_c = w(z_c)$ and plotted against $z/z_c$, with $z_c$ denoting the crossover altitude between the two regimes [Fig.~\ref{fig:Fig3}(a)]. Despite local latent heat exchanges associated with condensation and evaporation---which follow the spatial variations of $\rho_\ell$---this collapse simplifies the macroscopic description: turbulent mixing in the cloud behaves as if driven by a uniform effective buoyancy flux $B$. To quantify this flux, we analyze the trajectory of a cloud generated via horizontal injection [Fig.~\ref{fig:Fig3}(b)]. In agreement with a self-similar trajectory theory \cite{Note1}, the cloud centerline follows a cubic trajectory, $z \propto B x^3$. We predict and experimentally confirm that $B$ scales linearly with the injected mass flow rate $\dot{m}$. This buoyancy flux accounts for three distinct physical contributions: the temperature excess of the superheated vapor, the latent heat released during condensation, and the lower molar mass of water vapor compared to dry air, yielding $B = \beta g \dot{m}$ [Fig.~\ref{fig:Fig3}(c)]. The measured coefficient, $\beta = 5.3$, is slightly more than half of the theoretical upper bound, $\beta_{\max} \simeq 8.5$, obtained by summing these three contributions under full condensation. This difference is consistent with the partial condensation of the injected vapor.

This simplification allows for a mean-field model that neglects microphysical heterogeneities and finite phase-change relaxation times. This model enforces the conservation of fluxes for dry air (density $\rho_a$), total water mass ($\rho_\ell + \rho_v$), and total enthalpy $(\rho_a C_p^a + \rho_v C_p^v + \rho_\ell C_p^\ell) T - \rho_\ell \mathcal{L}$, where $C_p$ denotes the respective specific heat capacities and $\mathcal{L}$ is the latent heat of vaporization. Assuming identical turbulent mixing rates, governed by $w$ and $\sigma$, for droplets, water vapor, and enthalpy~\cite{birch_turbulent_1978}, any conserved passive scalar field $f$ obeys the following self-similar scaling relation:
\begin{equation}
\frac{f - f_\infty}{f_s - f_\infty} = \frac{R_s^2 U_s}{(\lambda z)^2 w(z)} \exp\left(-\frac{r^2}{2(\lambda z)^2}\right) \equiv \phi(r,z),
\label{formphi}
\end{equation}
where the dimensionless similarity variable $\phi(r,z)$, as defined above, applies to both the jet and plume regimes [Eqs.~\eqref{wjet} and \eqref{wplume}]. Note that neither $\rho_\ell$ nor $\rho_v$ individually follow this scaling law; only their sum does, owing to the kinematically coupled transport of the two phases at low Stokes numbers. The cloud boundary, defined by the onset of saturation conditions 
\begin{equation}
\rho_v = \rho_{\mathrm{sat}}(T) \quad \text{and} \quad \rho_\ell = 0,
\label{jetisocontour}
\end{equation}
corresponds to a constant isocontour $\phi = \Phi$. Expanding the saturation relation around the ambient state $\rho_v^\infty \simeq \rho_{\mathrm{sat}}(T_\infty)$ yields
\begin{equation}
\Phi \approx \frac{1 - \frac{\rho_v^\infty}{\rho_{\mathrm{sat}}(T_\infty)}}{\left(\frac{m_v \mathcal{L}}{k_B T_\infty} - 1\right)\left(1- \frac{\rho_s C_p^v T_s}{\rho_\infty C_p^\infty T_\infty}\right) + \frac{\rho_s}{\rho_{\mathrm{sat}}(T_\infty)}-1},
\label{eq:Phi}
\end{equation}
where $m_v$ is the mass of a water molecule, $k_B$ is the Boltzmann constant, and we have approximated the mixture's heat capacity per unit volume by that of the ambient air, $\rho_{a} C_p^a + \rho_{v} C_p^v \approx \rho_\infty C_p^\infty$, far from the outlet \cite{Note1}.

The predicted axial profile of $\rho_\ell$ is superimposed on the experimental measurements in Fig.~\ref{fig:Fig2}(a), demonstrating agreement above $z_d$. The mean-field model predicts complete cloud evaporation at an altitude $z=\ell$, where $\rho_\ell$ vanishes. Experimentally, isolated droplet clusters survive before turbulent stretching-folding events mix them with the subsaturated environment. This exponential decay, attributed to Poisson-distributed mixing events \cite{pumir_exponential_1991,shraiman_lagrangian_1994,shraiman_scalar_2000,villermaux_mixing_2003}, reflects the Lagrangian persistence of weakly strained fluid parcels, allowing isolated droplet swarms to survive over extended distances. The characteristic length scale for these stretching-folding events is set by the integral scale of the turbulence, which here scales with the local jet or plume width $\sigma$ at the onset of the heterogeneous decay phase. Figure~\ref{fig:Fig2}(d) shows a proportional relationship between the exponential decay length $\ell_e$ and $\sigma$. This exponential decay persists down to the measurement noise floor without discontinuity, precluding the determination of an absolute evaporation altitude for the final surviving droplet \cite{de_rivas_dense_2016,chong_extended_2021}. Exploiting the self-similarity of the profiles in the evaporation zone, we establish a robust experimental definition of the cloud length $\ell$ by linearly extrapolating the mean profile of $\rho_\ell(z)$ from its inflection point to zero [Fig.~\ref{fig:Fig2}(c)].

\begin{figure}[t!]
\includegraphics{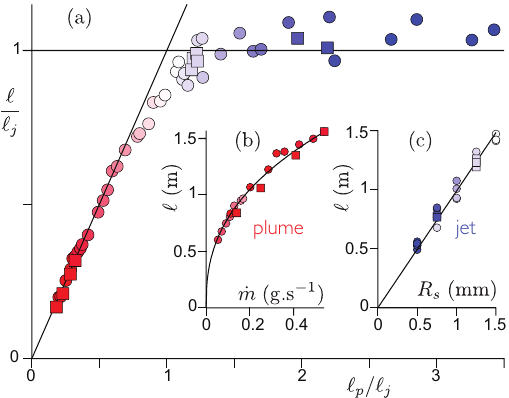}
\vspace{-4 mm}
\caption{(a) Rescaled cloud length $\ell/\ell_j$ as a function of the scaling ratio $\ell_p/\ell_j$.
(b) Cloud length $\ell$, corrected for humidity and temperature effects by a factor $(\Phi/\Phi_0)^{3/5}$ with $\Phi_0=2\times 10^{-2}$, as a function of the mass flow rate $\dot{m}$ in the plume regime. The solid black line denotes the theoretical prediction from Eq.~\eqref{ellplume}.
(c) Cloud length $\ell$, corrected for humidity and temperature effects by a factor $\Phi/\Phi_0$, as a function of the nozzle radius $R_s$ in the jet regime. The solid black line denotes the theoretical prediction from Eq.~\eqref{elljet}.
Squares (circles) represent experiments with (without) background incense smoke. The color encodes the ratio $\ell/z_c$, as in Fig.~\ref{fig:Fig2}(c).
}
\vspace{-4 mm}
\label{fig:Fig4}
\end{figure}

In the asymptotic model, we compute the cloud length $\ell$ from the condition $\phi(0,\ell) = \Phi$, which yields the following scaling laws:
\begin{align}
\ell_j &= \sqrt{\frac{\rho_\infty}{\rho_s}} \frac{R_s}{\lambda \Phi} \quad \text{(jet regime)}, \label{elljet} \\
\ell_p &= \left(\frac{2 \rho_\infty \dot{m}^2}{3 \pi^2 \beta \rho_s^3 \lambda^4 g \Phi^3}\right)^{1/5} \quad \text{(plume regime)}. \label{ellplume}
\end{align}
In the jet regime, $\ell_j$ scales linearly with the nozzle radius $R_s$ but remains independent of the mass flow rate $\dot{m}$ [Fig.~\ref{fig:Fig4}(c)], reflecting inertia-dominated dynamics. Conversely, in the plume regime, $\ell_p \propto \dot{m}^{2/5}$ and is independent of $R_s$ [Fig.~\ref{fig:Fig4}(b)], characteristic of buoyancy-driven entrainment. The dimensionless threshold $\Phi$ unifies the description of the cloud boundary across both flow regimes. As shown in Fig.~\ref{fig:Fig4}(a), plotting $\ell/\ell_j$ as a function of the scaling ratio $\ell_p/\ell_j$ collapses the experimental data onto a single master curve. Despite the model's mean-field approximations and the finite spatial extent of the transient condensation zone, both regimes are quantitatively predicted, exhibiting a plateau at $\ell/\ell_j \simeq 1$ in the jet limit and transitioning to the plume regime for $\ell_p/\ell_j \gtrsim 1$.

\textit{Conclusion}---This Letter extends the classical asymptotic analysis of turbulent jets and plumes to incorporate phase transitions, establishing a robust framework for describing the macroscopic evolution of droplet-laden clouds. The spatial extent $z_d$ of the near-injector transient condensation zone is governed by the competition between the microphysical relaxation time and the local turbulent advection time. Our experimental results confirm that higher CCN concentrations $\Psi$ accelerate the initial condensation phase, thereby significantly shortening $z_d$. Beyond this transient region, the assumption of interstitial thermodynamic equilibrium accurately captures the essential physics governing the cloud's spatial extent and persistence. This is strikingly validated by the collapse of experimental data onto a universal master curve spanning both the jet and plume regimes.

Our analysis further highlights the inverse dependence of the cloud boundary threshold $\Phi$ on ambient humidity [Eq.~\eqref{eq:Phi}], whereby higher moisture environments naturally enhance cloud persistence~\cite{chong_extended_2021}. The parameter $\beta$, encapsulating the latent heat released during condensation, emerges as a central driver of the buoyancy flux, effectively coupling local microphysics to the large-scale flow dynamics. While our asymptotic framework successfully filters out the complexities of droplet size distributions to provide robust first-order predictions, it also paves the way for future refinements. Extending this Letter via numerical modeling will help quantify latent heat-buoyancy feedbacks under complex humidity gradients, while further experimental studies could probe the prolonged survival of localized droplet clusters during the final heterogeneous evaporation phase. Ultimately, adapting these laboratory-derived scaling laws to atmospheric flows will test the saturated humidity assumption against field data, a crucial step for refining predictions of cloud behavior in both natural and industrial contexts.

\textit{Acknowledgments}---The authors thank B.~Amar, C.~No\^us, E.~Rolley, and B.~Semin for their insightful suggestions. This work was supported financially by ANR Apieds and ANR Mesclun.

\textit{Data availability}---All data generated or analyzed during this study are included in this published article and its Supplemental Material files \cite{Note1}.

\bibliographystyle{apsrev4-1}
%
\end{document}